\documentclass{optica-article}

\journal{oe}

\usepackage{lineno}
\usepackage{layouts}

\usepackage{siunitx}
\DeclareSIUnit{\bel}{B}

\newcommand{\AdevAbstract}{$  \sim 10^{-18}$ at $\SI{100}{\second}$ }

\newcommand{\samplingrate}{\SI{125}{\mega S \per \second}}

\newcommand{\AwgPartNumber}{\textit{Keysight 33500B}}
\newcommand{\vcopartnumber}{\textit{Minicircuits ZOS-75+}}
\newcommand{\AOMpartnumber}{\textit{Brimrose AMF-55-1550-2FP}}
\newcommand{\RFAMPpartnumber}{\textit{ZHL-1-2W+} for remote A and \textit{ZPUL-30p} for remote B}
\newcommand{\BPfilterPartNumber}{\textit{Minicircuits BBP-60+}}
\newcommand{\RBclockparthnumber}{\textit{SRS FS-725}}

\newcommand{\stemlabpartnumber}{\textit{STEMLAB 125-14}}
\newcommand{\SymetriconPartNumber}{\textit{Microsemi 3120A}}

\newcommand{\nhopping}{N_{\mathrm{hopping}}}
\newcommand{\nbits}{N}

\newcommand{\Disp}[1]{\left(\frac{\partial \varphi}{\partial \omega}\right)_{#1}}

\begin{document}
\title{Ultrastable optical frequency dissemination over a branching passive optical network using CDMA}

\author{Rodrigo Gonz\'alez Escudero,\authormark{1} Sougandh Kannoth Mavila,\authormark{1} 
and Jeroen C. J. Koelemeij \authormark{*,1}}

\address{\authormark{1}LaserLaB, Department of Physics and Astronomy, Vrije Universiteit Amsterdam, 1081 HV Amsterdam, Netherlands}

\email{\authormark{*}j.c.j.koelemeij@vu.nl} 



\begin{abstract}
We demonstrate a technique for ultrastable optical frequency dissemination in a branching passive optical network using code-division multiple access (CDMA). In our protocol, each network user employs a unique pseudo-random sequence to rapidly change the optical frequency among many distinct frequencies. After transmission through the optical network, each user correlates the received sequence with the transmitted one, thus establishing a frequency-hopping spread spectrum technique that helps reject optical signals transmitted by other users in the network. Our method, which builds on the work by Schediwy et al. [Opt. Lett. {\bf 38}, 2893 (2013)], improves the frequency distribution network's capacity, helps reject phase noise caused by intermediate optical back scattering, and simplifies the operational requirements. Using this protocol, we show that a frequency instability better than \AdevAbstract while having more than 100 users operating in the network should be possible. Finally, we theoretically explore the limits of this protocol and show that the demonstrated stability does not suffer from any fundamental limitation. In the future, the CDMA method presented here could be used in complex time-frequency distribution networks, allowing more users while, at the same time, reducing the network's complexity.
\end{abstract}
 
\section{Introduction}
Since its earliest demonstration~\cite{Maol1990}, ultrastable optical frequency dissemination using fiber-optic links has seen a great deal of development, including many demonstrations over long-haul links~\cite{WilliamsJOSAB2008,PredehlScience2012,DrostePrl2013,Schioppo2022natcoms} and the establishment of frequency distribution networks with complex topologies such as branches and loops~\cite{clonets,CantinNJP2021,HusmannOE2021,InseroScientificReports2017}. These developments are motivated by a plethora of applications such as quantum key distribution~\cite{ClivatiNature2022,LucamarininiNature2018}, seismic sensing~\cite{MarraScience2018, MarraScience2022}, precision spectroscopy for the determination of fundamental constants~\cite{MatveevPRL2013}, and relativistic geodesy ~\cite{ChouScience2010,GrottiNatPhys2018,TakamotoNatPhot2020}.

To fully realize these goals it is desirable to deliver the optical signal to as many users as possible while keeping the network's complexity and cost low. In most of the existing implementations, this target is hindered by the transmission technology used. For example, multi-user branching networks are formed by creating parallel point-to-point connections, requiring involved equipment including repeater lasers at each branching site~\cite{CantinNJP2021}. In addition, the branching device may have to be located at intermediate sites with limited access (such as data centers), where installing (additional) hardware might be relatively complicated and costly.

Some of the aforementioned challenges have been resolved in part using automated and standardized systems~\cite{Guillou-CamargoAppliedOptics2018}. Another method for cost-effective distribution of an ultrastable optical carrier signal along a linear stabilized link was demonstrated in Refs.~\cite{GroscheOL2014,BaiOL2013}. A particularly advantageous branched network for ultrastable frequency dissemination was demonstrated in Ref.~\cite{SchediwyOE2010}. Here, a single ultrastable signal was distributed over several fiber-optic branches using a simple passive optical power splitter, which makes the method inherently compatible with relatively inexpensive passive optical networks. At the end of each branch, an electro-optical unit takes care of the stabilization of the optical path between the ultrastable laser source and the end of the branch, and simultaneously provides the ultrastable optical signal to the local user. Here, the detection of fiber-length variations is based on a beat note between the original ultrastable carrier and a frequency-shifted optical wave, produced by the electro-optical unit, that makes a round trip from the branch's end to the ultrastable laser source and back. The latter wave therefore passes through the optical splitter twice, which inevitably leads to cross talk between different user sites. Schediwy et al. solve this by assigning slightly different frequency shifts to the round-trip wave of each user, so that after detection electrical bandpass filters can be used to reject the undesired signals from other users in the network\cite{SchediwyOE2010}. Therefore each user must use a unique optical frequency, which requires a certain amount of coordination among the users themselves or by a central administrator. Because link stabilization requires some bandwidth around the optical frequency of each user, the number of users is ultimately limited. Another potential weakness is the sensitivity to rogue users or attackers, which could (un)intentionally disable an arbitrary number of branches by transmitting in frequency channels that had already been assigned to other users.

The technique presented in Ref.~\cite{SchediwyOE2010} is a form of frequency-division multiple access (FDMA). FDMA was used in early generations of mobile networks and its limitations are well known \cite{GramiBook2016}. By contrast, later generations improved this implementation using code-division multiple access (CDMA). In CDMA, each receiver uses a unique pseudo-random sequence and correlates the received signal with the expected sequence to reject signals of other users in the network.

Here we show that CDMA can be extended to the realm of ultrastable optical frequency dissemination. In our implementation, the pseudo-random sequence is encoded in the optical signal by hopping over many distinct optical frequencies. After transmission through the optical system, we recover the original sequence and reject the effect of other users in the network. We show that CDMA can be used to transfer the optical reference while having many users operating in the network and we explore the limitations of this dissemination technique. We estimate that a fractional frequency stability of \AdevAbstract while having more than 100 users operating in the network is possible. We explore the limitations of this technique and show that with further improvements more users are possible. Finally, our method rejects reflections inside the fiber, a common source of noise \cite{WilliamsJOSAB2008}, reducing the number of instruments required inside an optical fiber link. The CDMA method presented here could ease the development of complex (passive) optical networks for ultrastable frequency dissemination, by allowing many more users to operate in the same network while reducing the hardware requirements. 

We note that our approach is similar in spirit to the technique described in \cite{Jochen_pattent}; however, to the best of our knowledge, our work represents the first demonstration of frequency-hopping CDMA for optical link stabilization and frequency transfer. We also point out that cross-correlation of frequency-hopping sequences was recently used to transfer time over fiber links up to 150~km in length with picosecond stability, while observing similar suppression of effects of unwanted reflections~\cite{XieTUFFC2024}. Cross-correlation of 10~Gb/s on-off-keyed optical data, for the purpose of fiber-optic delay measurements over 75~km distance with few picoseconds uncertainty, was demonstrated earlier by Sotiropoulos et al.~\cite{Sotiropoulos2013}.

\section{Protocol description}

\subsection{CDMA single link}
\label{subsec:single_link}

\begin{figure}[t!]
\centering\includegraphics[width= \textwidth]{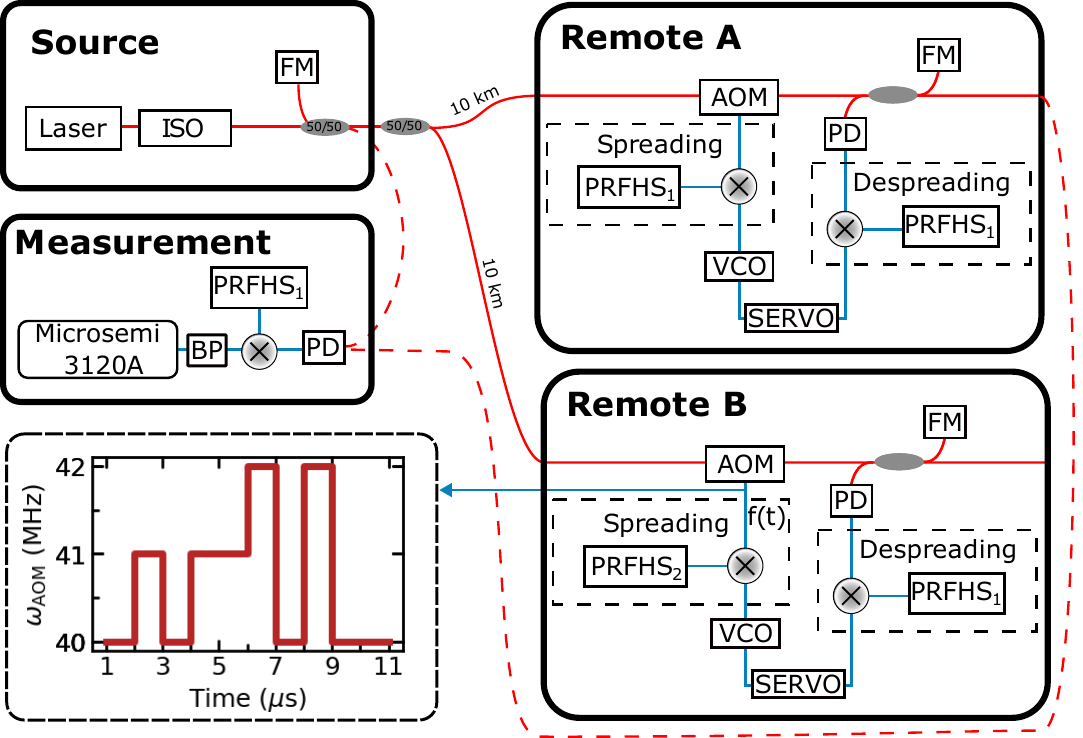}
\caption{Simplified diagram of the experimental setup. A more detailed description of each remote site including all relevant experimental components used is shown in figure \ref{fig:nodedetailed}. The dashed box shows a possible frequency hopping sequence used to drive the AOM. The shown sequence follows equation \ref{eq:f_1} with $f_{0} = \SI{40}{\mega\hertz}$, $\Delta\!f = \SI{1}{\mega\hertz}$, $t_{s} = \SI{1}{\micro \second}$ and ${\nhopping} = 3$. FM, Faraday mirror; PD, photodetector; VCO, voltage-controlled oscillator; PRFHS, pseudorandom frequency hopping sequence; ISO, optical isolator; BP, bandpass filter. }
\label{fig1:setup}
\end{figure}

Figure \ref{fig1:setup} displays several branches of a network for ultrastable frequency dissemination using CDMA. To provide insight into our protocol's principle of operation, we follow an optical signal traveling from the signal source to remote location A. For clarity, we will ignore
optical signals associated with any other links. A more general treatment will be given in subsection~\ref{subsec:multiple_links}.



The signal from an ultrastable laser source is sent through an optical fiber link of length $L$ ($\SI{10}{\kilo\meter}$ in figure~\ref{fig1:setup}). A key property of the ultrastable laser is that its coherence length well exceeds the typical length scales in the fiber-optic network. During transmission through the optical fiber, random optical path length variations lead to a
time-dependent phase shift, $\delta\!\phi (t)$, given by~\cite{PinkertApplOpt2015}:
\begin{equation}
\delta\!\phi (t) = 2\pi\times\frac{\nu_{0}}{c} \left[ n(\nu_{0},t) L(t) \right].
\end{equation}

Here, $\nu_{0}$ is the optical frequency of the laser, $n$ is the index of refraction of the medium, and $c$ is the speed of light in the medium. In principle, $\delta\!\phi (t)$ varies over time with a power spectrum that covers a wide range of frequencies~\cite{WilliamsJOSAB2008}. However, we only consider link-induced phase shifts that lead to quasi-static frequency offsets; i.e, on the time scale of the one-way propagation delay through the fiber, the corresponding optical path length changes (or ’noise’) are approximately linear.

At the remote site, the optical signal passes through an acousto-optic modulator (AOM) at time $t$, and its time-dependent phase becomes:
\begin{equation}
\label{eq_nuA1}
\phi_\mathrm{A1}(t) = 2\pi t  \times \left[\nu_{0} + f_{\mathrm{s}}(t)\right] + \delta\!\phi (t).
\end{equation}
Here the additional time dependence in the frequency of the AOM, $f_{\mathrm{s}}(t)$, is introduced to accommodate the frequency-hopping sequence used in our protocol. In the parlance of CDMA, $f_{\mathrm{s}}(t)$ serves to 'spread' the optical signal.

A portion of the optical signal is reflected back towards the signal source by a Faraday mirror. After passing through the AOM and the fiber link, the phase of this optical signal becomes $\phi_\mathrm{A2}(t+\tau) =2\pi t \times\left[ \nu_{0} + 2f_{\mathrm{s}}(t)\right]+ 2\delta\!\phi (t)$, where $\tau = L/c$ is the propagation delay of the light in the fiber. Here we neglect the small (few nanosecond) propagation delay between the AOM and the photodetector, in view of the much larger propagation delays and timing uncertainties in the system.

Finally, after reflection by the Faraday mirror at the source and the subsequent transmission through the fiber link, the optical signal arrives at remote location A at time $t+2\tau$, where it undergoes one additional pass through the AOM before it is detected. The phase thus becomes

\begin{equation}
\label{eq_nuA3}
\phi_\mathrm{A3}(t+2\tau) =2\pi t \times \left[\nu_{0} + 2f_{\mathrm{s}}(t)  + f_{\mathrm{s}}(t + 2\tau)\right] + 3\delta\!\phi (t) ,
\end{equation}

where the term $3\delta\!\phi (t)$ represents the phase shift due to link noise after propagating through the fiber link three times.

Now, consider the beat note measured at the photo-detector at time $t$. With the help of Eqs.~(\ref{eq_nuA1}) and (\ref{eq_nuA3}), this can be written as:

\begin{equation}
\label{eq:fbeat}
\phi_{\mathrm{beat}}(t) = \phi_\mathrm{A3}(t) - \phi_\mathrm{A1}(t) = 2 \pi t \times 2f_{\mathrm{s}}(t-2\tau) +2\delta\!\phi (t) ,
\end{equation}

The beat-note signal is mixed with a time-dependent electrical signal, $\phi_{\mathrm{d}}(t)$ which serves to 'despread' the detected signal (as will be explained below). The low-pass filtered signal produced by the mixer product becomes:

\begin{equation}
\label{eq:Smixer_1}
\phi_\mathrm{mixer}(t) =\phi_{\mathrm{beat}}(t) - \phi_{\mathrm{d}}(t) = 2 \pi t \times 2f_{\mathrm{s}}(t-2\tau) +  2\delta\!\phi (t) - \phi_{\mathrm{d}}(t),
\end{equation}
The mixer product is subsequently used as an error signal for a controller, which acts on the VCO that steers the AOM. This controller thus produces a control frequency $\delta\!f$, which is used to compensate the link noise $\delta\!\phi (t)$ in the optical domain. Note that when the link stabilization is active, we nominally expect to have $\phi_\mathrm{mixer}(t)= 0$.

Before feeding the control frequency into the AOM, it is mixed with the spreading signal, $f_{\mathrm{s}}(t)$ [Eq.~(\ref{eq:f_1})]. As a result, the frequency of the AOM jumps between $\nhopping$ possible frequencies (or 'symbols'), the order of which is chosen according to a pseudo-random sequence, $s(t)$, that is unique to each user. The duration of each frequency symbol is $t_{s}$. An example sequence is shown in the dashed box of Fig.~\ref{fig1:setup}. The frequency sent into the AOM can be expressed mathematically as follows:

\begin{equation}
\label{eq:f_1}
f_{\mathrm{s}}(t) = f_{0} + \Delta\!f \cdot s(t) + \delta\!f.
\end{equation}
Here, $\Delta\!f$ is the spacing between adjacent frequency levels, and $f_{0}$ is an overall frequency offset.
Inserting Eq.~(\ref{eq:f_1}) into Eq.~(\ref{eq:Smixer_1}) and choosing $\phi_{\mathrm{d}}(t)$ such that $\phi_{\mathrm{d}}(t) = 2\pi  t \times \left[2(f_{0} + \Delta\!f \cdot s(t-2\tau_{\mathrm{d}}))\right]$, with $\tau_{\mathrm{d}}$ a programmable delay, we obtain:

\begin{equation}
\label{eq:Smixer}
\phi_\mathrm{mixer}(t) = 2\pi t \times \left[2\Delta\!f(s(t-2\tau) - s(t-2\tau_{\mathrm{d}}))\right] + 2\left[2 \pi t \delta\!f + \delta\!\phi(t)\right].
\end{equation}

Crucially, by setting $\tau_{\mathrm{d}}$ equal to the fiber propagation delay $\tau$, the above equation reduces to $\phi_\mathrm{mixer}(t) = 2\left[2 \pi t \delta\!f + \delta\!\phi(t)\right]$. Conversely, if $\tau_{\mathrm{d}}$ is chosen incorrectly, the mixer output signal becomes modulated at frequencies that are (sub)multiples of $1/t_s$, which may be higher than the servo control bandwidth. Finally, enabling the servo loop will enforce the condition $\phi_\mathrm{mixer}(t)=0$ so that $2 \pi t \delta\!f = -\delta\!\phi(t)$, thus compensating the noise induced by the link.

The frequency-hopping scheme implies propagation delay differences due to chromatic dispersion, which may lead to significant phase jumps in the compensated optical signal. From the material index of refraction of silica optical fiber~\cite{Ghosh1994}, and for an optical wavelength of $\SI{1.5}{\micro \meter}$, we coarsely estimate a  differential phase delay of about $\SI{40}{\pico \second \per \nano \meter \per \kilo \meter}$ (a more accurate estimate would consider also the waveguide dispersion). For frequency jumps of about $\SI{1}{\mega \hertz}$, for example, the differential delays are about $\SI{0.03}{\pico \second}$ (assuming a link length of $\SI{100}{\kilo \meter}$). Even though these delays are negligible compared to other relevant time scales in the experiment, they induce optical phase jumps larger than $2\pi$. Consequently, the optical phase of the signal transmitted through the fiber will jump in a pseudo-random fashion every $t_{s}$ seconds.

For small frequency jumps such that the dispersion can be considered constant, these phase jumps can be expressed in terms of the PRFHS as $\frac{\partial \varphi}{ \partial \omega} \cdot 2\pi\Delta\!f \cdot s(t)$. Here, $\frac{\partial \varphi}{\partial \omega}$ is the optical phase delay per unit angular frequency, for a fiber with length $L$ and light with a given optical wavelength, $\lambda$.

In the following, we modify the previous equations to account for this phase modulation. First, the signal presented in Eq.~(\ref{eq:fbeat}) will carry this additional modulation as follows:
\begin{equation}
    \phi_{\mathrm{beat}}(t) = 2 \pi t \times 2f_{\mathrm{s}}(t-2\tau) +2\delta\!\phi (t) + \frac{\partial \varphi}{ \partial \omega} \cdot 2\pi\Delta\!f \cdot s(t-2\tau).
\end{equation}
To compensate for the dispersion-induced term in the beat signal, we add an additional phase modulation $\Delta \varphi\cdot s(t-2\tau_{\mathrm{d}})$ to the despreading signal, which becomes:
\begin{equation}
\phi_{\mathrm{d}}(t) = 2\pi  t \times \left[2(f_{0} + \Delta\!f \cdot s(t-2\tau_{\mathrm{d}}))\right] + \Delta \varphi\cdot s(t-2\tau_{\mathrm{d}})
\end{equation}
%
%
%
%

Finally, following a similar algebra as above for Eq.~(\ref{eq:Smixer}), we arrive at a mixer output signal with phase:
\begin{eqnarray}
\label{eq:Smixer_With_phase_delays}
\phi_\mathrm{mixer}(t) &=& 2\pi t \times \left[2\Delta\!f(s(t-2\tau) - s(t-2\tau_{\mathrm{d}}))\right]  + 2\left[2 \pi t \delta\!f + \delta\!\phi(t)\right] \nonumber \\  &+& \frac{\partial \varphi}{ \partial \omega} \cdot 2\pi\Delta\!f \cdot s(t-2\tau) - \Delta \varphi\cdot s(t-2\tau_{\mathrm{d}}) .
\end{eqnarray}
By setting $\tau_{\mathrm{d}} = \tau $ and $ \Delta \varphi = \frac{\partial \varphi}{ \partial \omega} \cdot 2\pi\Delta\!f  $, a cancellation of terms in Eq.~(\ref{eq:Smixer_With_phase_delays}) occurs such that $\phi_\mathrm{mixer}(t) =  \left[2 \pi t \delta\!f + \delta\!\phi(t)\right]$, which can be used to cancel the link noise as before. In the supplementary materials (section \ref{s:FDC})  we describe how the above conditions are achieved in practice.

\subsection{CDMA multiple links}
\label{subsec:multiple_links}

The description of the previous subsection does not consider the effect of any additional links. However, light from the signal source can reflect off a second remote location B (meanwhile traversing the AOM at location B twice), and subsequently reflect off the signal source to finally reach remote location A. This additional optical signal has the potential to contaminate the output of the photodetector at location A, PD$_{\mathrm{A}}$, with
the control signal of remote location B as well as the noise of the link connecting to remote location B. As we demonstrate below, the spreading at location A modulates this undesired signal to frequencies that are mostly outside the passband of the filter in front of controller A (Fig.~\ref{fig:nodedetailed}), and outside of the phase-locked loop (PLL) bandwidth. However, some cross talk does occur, and we quantitatively assess its impact here.

The evaluation of cross talk starts out from the interference between the paths of links A and B at PD$_{\mathrm{A}}$, which involves three optical fields which we will label as $\phi_\mathrm{A1}$, $\phi_\mathrm{A3}$, and $\phi_\mathrm{A1B2}$. As before, the numerical index indicates the number of one-way passes through each link, starting from the ultrastable laser source. The resulting beat notes are mixed with the despreading signal and low-pass filtered, leading to a signal sent to the controller that contains the following two terms:
\begin{equation}
\label{eq:fourterms}
\cos \left(\phi _{\text{A1B2}} - \phi _{\text{A1}}  - \phi _{\text{d}}\right)+\cos \left(\phi _{\text{A3}} -\phi _{\text{A1}}-\phi_\mathrm{d}\right)
\end{equation}
In principle, multiple reflections between the Faraday mirrors will result in other terms that should be included in the previous equation. For example, a beat note between $\phi_\mathrm{A3}$ [Eq.~(\ref{eq_nuA3})] and $\phi_\mathrm{A1B2}$ [Eq.~\ref{eq_nuA1B2})] is possible. These additional terms require a larger number of total passes through the optical system (six in this particular example) resulting in a much stronger attenuation. Therefore, we do not include them here.

Following a similar line of reasoning as before, the phase of the light from remote location B reaching the photodetector at A can be calculated to be:
\begin{eqnarray}
\label{eq_nuA1B2}
\phi_\mathrm{A1B2}(t) &=& 2\pi t \times \left[\nu_{0} + 2f_\mathrm{s,B}(t - \tau_\mathrm{A} - \tau_\mathrm{B}) + f_\mathrm{s,A}(t) \right] + 2\delta\!\phi_{\mathrm{B}} (t) + \delta\!\phi_{\mathrm{A}} (t)\nonumber \\&+& 2\pi \Disp{\mathrm{AB}} \Delta f \cdot s_\mathrm{B}(t-\tau_\mathrm{A}-\tau_\mathrm{B}) 
\end{eqnarray}



Here, we have introduced the additional subscripts A and B to refer to the time delays, PRFHS, and fiber noise associated with links A and B, respectively. Note also that the factor $(\partial \varphi / \partial \omega)_\mathrm{AB}$ accounts for the total chromatic dispersion sustained during the subsequent transmission through optical paths B and A.

The signal given in Eq.~(\ref{eq_nuA1B2}) will interfere with $\phi_\mathrm{A1}$ (Eq.~\ref{eq_nuA1}) to produce an additional beat note at the photodiode of remote A of:
\begin{eqnarray}
\label{eq_nuA21}
 &&\phi_\mathrm{A1B2}(t) - \phi_\mathrm{A1}(t) = \\  &&2\pi t \times  2f_\mathrm{s,B}(t - \tau_\mathrm{A} - \tau_\mathrm{B}) + 2\delta\!\phi_{\mathrm{B}} (t) + 2\pi \Disp{\mathrm{AB}} \Delta f \cdot s_\mathrm{B}(t-\tau_\mathrm{A}-\tau_\mathrm{B}) 
\end{eqnarray}

%
Therefore, the low-pass filtered mixer product at remote location A will pick up the additional signal with phase
\begin{eqnarray}
\label{eq:SmixerB}
\phi_\mathrm{mixer,AB} & = & \phi_\mathrm{A1B2}(t) - \phi_\mathrm{A1}(t) - \phi_\mathrm{d,A} \nonumber \\
&=& 2\pi t \times \left[2\Delta\!f \left[ s_\mathrm{B}(t - \tau_\mathrm{A} - \tau_\mathrm{B}) - s_\mathrm{A}(t-2\tau_{\mathrm{d,\mathrm{A}}})\right] + \delta\!f_{\mathrm{B}}\right] + 2\delta\!\phi_{\mathrm{B}} (t)\\
&&\nonumber + 2\pi \Disp{\mathrm{AB}} \Delta f \cdot s_\mathrm{B}(t-\tau_\mathrm{A}-\tau_\mathrm{B}) -\Delta \varphi\cdot s_\mathrm{A} (t-2\tau_{\mathrm{d,A}})  + \varphi_\mathrm{slow}
\end{eqnarray}


Here, the subscript AB in $\phi_\mathrm{mixer,AB}$ indicates that this frequency is associated with the signal produced by mixer A originating from interference with the signal from remote location B. We furthermore assume that $\Delta\!f$
and $f_{0}$ are the same for both locations A and B. A more spectrally efficient situation where each user employs different values $f_\mathrm{0,A}, f_\mathrm{0,B}, \dots$ is not considered here~\cite{SchediwyOE2010}. 

Finally, we have introduced the slowly varying (such that it can be treated as a constant) phase factor $\varphi_\mathrm{slow}$. This phase arises from different optical and electronic time delays in each of the branches, the phase difference between each of the radio frequency sources used to generate the frequency hopping sequence, and the short yet uncompensated fiber sections in each of the branches. 
%


\subsection{Signal spectrum}
To analyze the suppression of optical signals arriving from other nodes, we calculate the power spectrum of the time-dependent signal given by Eq.~(\ref{eq:SmixerB}). The derivations of the equations presented here can be found in the supplementary materials (section \ref{s:FHSSdev}). Assuming unit signal amplitude, the power spectrum of this signal is given by $S_\mathrm{mixer,\mathrm{AB}}(\omega)$, with
\begin{multline}
 \label{eq:PSD}
S_\mathrm{mixer,\mathrm{AB}}(\omega) = \left | \frac{1}{T} \int_{0}^{T} \sin{(\phi_\mathrm{mixer,AB}(t))}e^{i\omega t} dt\right|^2 = \\ \left |\frac{1}{2T}\sum_{n=1}^{\nbits} e^{i \phi_{n}} \frac{e^{i (\omega_{n} - \omega_{k})n t_{s}} - e^{i(\omega_{n} - \omega_{k})(n+1) t_{s}} }{\omega_{n} - \omega_{k}} \right.  \left. +\; e^{-i\phi_{n}}\frac{e^{-i (\omega_{n} + \omega_{k})n t_{s}} - e^{-i (\omega_{n} + \omega_{k})(n+1) t_{s}} }{\omega_{n} + \omega_{k}} \right| ^2 ,
\end{multline}
where $\omega_{n} = 4 \pi\,\Delta\!f\!\,m_{n}$ is the angular frequency difference between PRFHS~A and PRFHS~B at symbol $n$, $m_{n}$ is an integer in the range $ \left[-m, m\right]$, $\nbits$ is the total number of pseudo-random symbols after which the signal repeats itself, $t_{s}$ is the duration of each bit, and $\phi_{n}$ is a phase value which is constant on the interval $\left[n t_{s}, (n + 1) t_{s}\right]$ . 
The low frequency components of Eq.~(\ref{eq:PSD}) become:
\begin{equation}
\label{fourier_serie3s}
S_\mathrm{mixer,\mathrm{A,B}}(0) = \left |\frac{1}{T}\sum_{n=1}^{\nbits} \frac{\cos({\omega_{n} n t_{s}}  +\phi_{n})-\cos({\omega_{n} (n+1)t_{s}} +\phi_{n})}{\omega_{n}} \right|^{2}.
\end{equation}
Crucially, at low frequencies ($\omega \ll f_{n}$) and assuming that the number of bits ($N$) is much larger than the number of hopping levels ($N_\mathrm{hopping}$), 
Eq.~(\ref{eq:PSD}) can be approximated as:
\begin{equation}
 \label{eq:PSD_0}
S_\mathrm{mixer,\mathrm{A,B}}(0) \approx \left |\frac{\sum_{n = 1}^{\nhopping}\sin{\phi^{\prime}_{n,\mathrm{unique}}}}{ \nhopping ^2} \right|^2.
\end{equation}
Here $\phi^{\prime}_{n,\mathrm{unique}}$ indicates that we are summing over the $\nhopping$ unique $\phi_{n}$ values where both branches A and B have the same hopping frequency.

This DC component is an undesired signal leaking from remote B that falls within the servo bandwidth of remote A and cannot be filtered out. It could result in cross-talk between each of the PLLs, or phase noise leaking from other users. Fortunately, as we will demonstrate below, this signal can be arbitrarily suppressed with a suitable number of hopping frequencies. 

\section{Experimental demonstration}
\begin{figure}[t!]
\centering\includegraphics[width= 7cm]{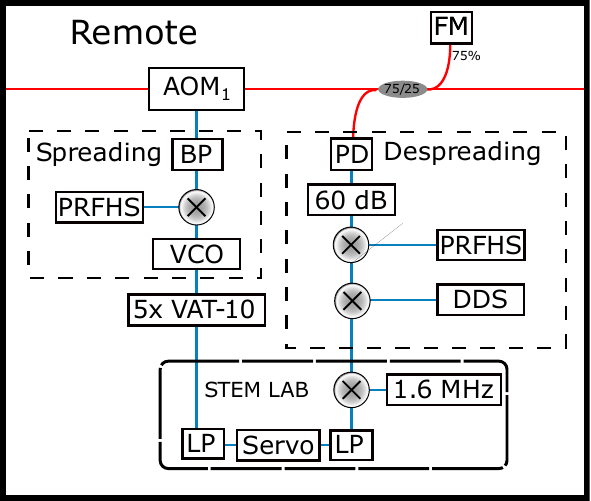}
\caption{Detailed diagram of the components used in each remote location.  BP, bandpass filter; LP, lowpass filter }
\label{fig:nodedetailed}
\end{figure}

Figure \ref{fig1:setup} shows a simplified schematic of our experimental setup. For clarity, this figure includes only those key components required to understand our experimental technique. A detailed description of each remote site, including every optical and electrical component used is shown in figure \ref{fig:nodedetailed}. 
We demonstrate our technique using two independent branches, A and B. The following description is common for both of them. The frequency hopping sequence is generated using an arbitrary waveform generator (AWG,~\AwgPartNumber). The AWG has two channels, which we use for spreading and despreading the signal. First, the output of a VCO (\vcopartnumber) is mixed with one of the channels of the AWG to produce the frequency hopping signal which, after filtering (\BPfilterPartNumber) and amplification (\RFAMPpartnumber) is fed directly into an AOM (\AOMpartnumber).
%
%

Then, at the photodetector, we recover a frequency hopping signal that follows Eqs.~(\ref{eq:fbeat}) and (\ref{eq:f_1}). The output of the photodetector is then mixed with the other channel of the AWG to despread the signal. Due to the limited sampling rate of the AWG (\samplingrate) we employ an additional mixer driven by a direct digital synthesizer (DDS) to down-convert the despreaded signal to a frequency of about $\SI{1.6}{\mega \hertz}$. 
Finally, we use a digital servo controller (\stemlabpartnumber) with the PyRPL software \cite{pyrpl} to filter and mix down to DC the $\SI{1.6}{\mega \hertz}$ signal. The servo controller then acts on this down-converted and filtered signal to drive the VCO and compensate for the phase noise. The additional mixing down stage is done for convenience, to ease the characterization and operation of the PLL.

As shown in Fig.~\ref{fig1:setup}, the output of remote A is used for an out-of-loop characterization of the frequency stability from the link. Note that the frequency hopping sequence is also imprinted in the output, so we have to despread the signal using an additional AWG. The resulting frequency is then bandpass filtered and down-converted to about $\SI{13}{\mega \hertz}$ using an additional mixer. Finally, the relative frequency stability is analyzed using a phase noise probe (\SymetriconPartNumber).

Small frequency offsets between both branches could lead to additional noise suppression not originating from our CDMA spread-spectrum technique. While this could be an interesting avenue to explore, we exclude such effects to assess the performance of the CDMA technique alone as follows. First, we keep both hardware and software used in both branches as identical as possible (a small difference being the amplifiers used). Second, we use the same external clock (\RBclockparthnumber) for all frequency generators. We thus ensure that the frequencies for the phase locks of both branches are the same, and that both remote A and B hop over the same frequency values. This latter point is also verified experimentally by measuring the beat frequency of the out-of-loop beat frequency for both remote sites A and B.
\begin{figure}[t!]
\centering\includegraphics[width= \textwidth]{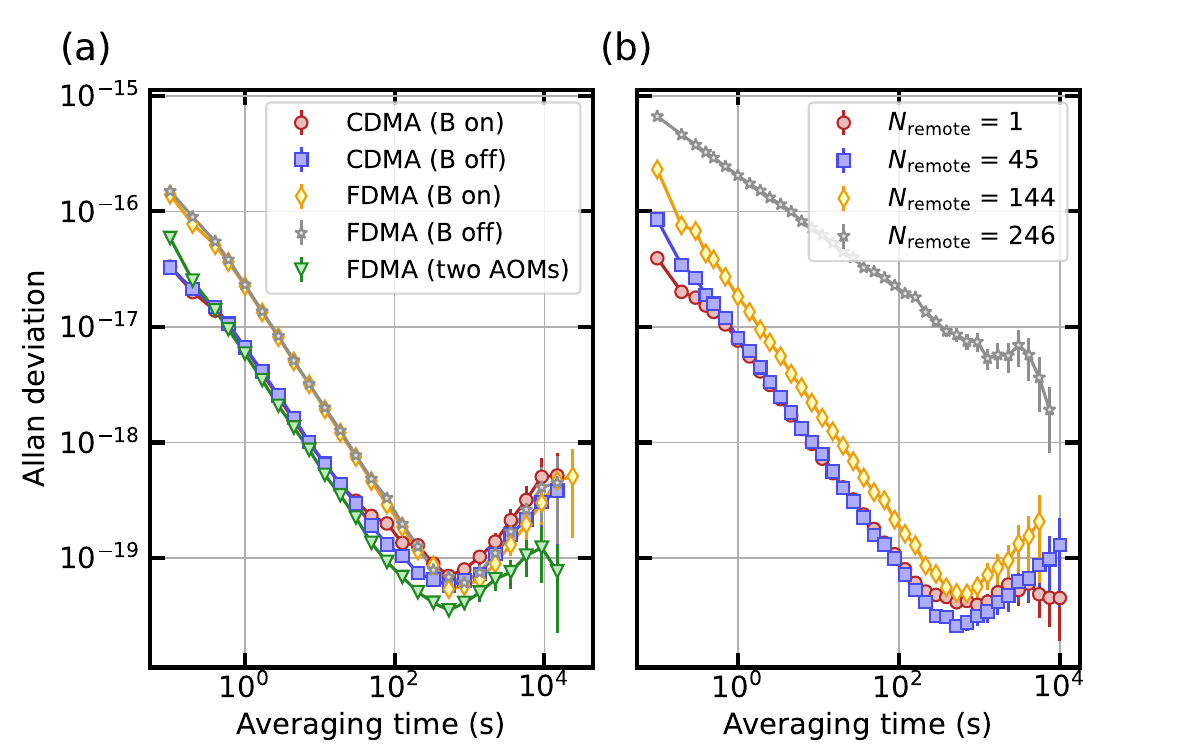}
\caption{ (a) Fractional frequency stability with remote B connected and disconnected using the CDMA protocol presented in this paper (red circles and blue squares respectively). For comparison we show the equivalent data using the FDMA protocol presented in Ref.~\cite{SchediwyOE2010} (yellow diamonds and grey stars); where both branches operate with a single frequency and a small frequency detuning. The green triangles show the frequency stability when an additional AOM is included. (b) Estimation of the maximum number of independent remote locations ($N_\mathrm{{remote}}$) that are possible using the CDMA protocol. }
\label{fig:adev}
\end{figure}
%
%

In Fig. ~\ref{fig:adev}~(a) we plot the relative frequency stability of the out-of-loop beat note, with remote B connected (red circles) or disconnected (blue squares). Our PRFHS is defined by $\nhopping= 31$, $t_{s} = \SI{2}{\micro \second}$, $\Delta\!f = \SI{100}{\kilo \hertz}$ and $\nbits = 100$ for remote A and $\nbits = 101$ for remote B. The slight difference between the number of bits is chosen to ensure that the probability that we reach a sequence where there is no frequency overlap between the two sequences at remote A and B is close to zero. With these parameters, the CDMA protocol shows a relative frequency stability lower than $10^{-17}$ at one second of averaging. More importantly, the inclusion of an additional user has a negligible effect on the transfer stability.

The grey stars and yellow diamonds of Fig.~\ref{fig:adev}~(a) show the stability when the frequency hopping sequence is disabled (which corresponds to the FDMA method introduced in Ref.~\cite{SchediwyOE2010}). Here we observe that the CDMA protocol delivers a slight performance improvement. We attribute this improvement to the suppression of reflections inside the fiber. Indeed, optical reflections within the link act as an additional source of noise~\cite{WilliamsJOSAB2008}, and recent work pointed out that spread spectrum techniques can suppress them~\cite{Jochen_pattent,XieTUFFC2024}. To confirm this behavior, we carried out measurements with an additional AOM included at the source site (green triangles). With this additional AOM, the reflections arrive at the photodetector at a different frequency and are thus rejected. When these reflections are eliminated, both methods show remarkably similar performance, confirming our hypothesis. The slightly better stability obtained by the additional AOM seems to indicate that our CDMA protocol does not fully suppress the optical reflections.


In Fig.~\ref{fig:adev} (b) we estimate the maximum number of independent remote locations ($N_\mathrm{{remote}}$) that are possible using our protocol. Here we use a different PRFHS than before with $\nhopping= 961$, $t_{s} = \SI{2}{\micro \second}$, of $\Delta\!f = \SI{3.125}{\kilo \hertz}$ and $\nbits = 3000$ for remote A and $\nbits = 3001$ for remote B. Note that due to limited availability of hardware in our laboratory, we were unable to demonstrate more than two independent remote sites, so to perform this estimation we include an erbium-doped fiber amplifier (EDFA) between the source and remote B. The number of sites is then deemed equal to the square of the amplifier gain. For this demonstration, we ensure that we operate the EDFA in the low-gain regime such that this approximation is correct. We verify this behavior by measuring the relative beat note strength at remote A between the signals originating from the two branches and comparing it to the measured amplifier gain.

For an EDFA gain corresponding to 45 locations, the achieved stability (Allan deviation) compares well with that achieved for a single-location network, except for a slight degradation of stability at averaging times below 1~s [see Fig.~\ref{fig:adev} (b)]. For EDFA gain corresponding to 144 locations, the stability increases by about a factor of two. We attribute this degradation of stability to the limited suppression of the noise originating from other branches. 

When increasing the EDFA gain further, to a corresponding number of 246 locations, the stability is degraded by over one order of magnitude at short averaging times. Moreover, the slope of the Allan deviation as a function of averaging time ($\tau_{av}$) is observed to shift from $\tau_{av}^{-1}$ to $\tau_{av}^{-1/2}$, indicating that a white frequency noise process dominates over the usual $\tau_{av}^{-1}$ delay unsuppressed noise~\cite{WilliamsJOSAB2008}. This behavior is caused by cycle slips that occur at the high EDFA gain level used. For implementations where this limit is not acceptable, a straightforward solution might be to increase the number of hopping levels spanning a larger frequency range. Introducing frequency offsets between different groups of users might also be an option. It should also be noted that in this demonstration, the fiber noise contribution of each remote site is considered to add linearly, i.e. $\propto {N_\mathrm{{remote}}}$, while in a real network a more reasonable scaling might be $\propto {N_\mathrm{remote}}^{1/2}$. 

From the data presented in Fig.~\ref{fig:adev} (b) we conclude that using our frequency-hopping CDMA protocol, relative frequency stability well below $10^{-18}$ at averaging times $\SI{100}{\second}$ in a network containing 144 sites should be possible, sufficient to distribute the frequency of state-of-the-art optical clocks~\cite{TakamotoNatPhot2020}. 
\begin{figure}[t!]
\centering\includegraphics[width= \textwidth]{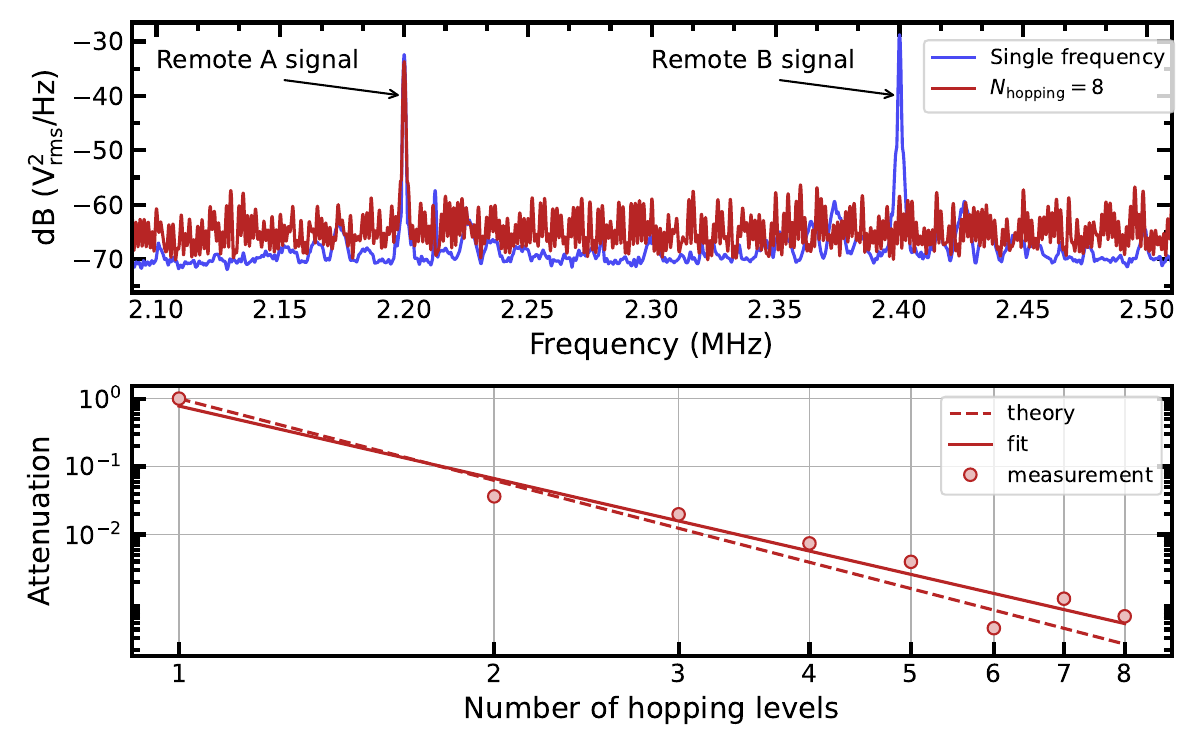}
\caption{(Upper panel) Power spectral density measured after the second mixing stage before filtering. The data compares to situations, when no modulation is applied (blue lines) and when the modulation is turned on (red lines). Here the PRFHS follows $t_{s} = \SI{2}{\micro \second}$, $ \nhopping = 961$, $ \Delta\!f = \SI{3.125}{\kilo \hertz}$ and $\nbits = 3000$. The data shows a suppression of $\sim \SI{30}{\deci\bel}$ of the signal from remote B. The spectrum is measured in remote A, at the input of the digital servo controller. (Lower panel)
Strength of the attenuation of the signal resulting from remote B. Shown are a least-squares fit to the data (dashed line) as well as the expected 1/$\nhopping^4$ dependence, derived theoretically in this work [Eq.~\ref{eq:PSD_0}].}
\label{fig:spectrum}
\end{figure}

To further illustrate the effect of spreading and despreading the frequency, we measure the power spectral density of the despread signal and compare it to the case where no modulation is applied.  In both cases, the data is measured at the input of the digital servo controller and before the last digital mixing stage (see Fig.~\ref{fig:nodedetailed}). The resulting data is shown in Fig.~\ref{fig:spectrum}. To unambiguously identify the signal resulting from each of the remote locations, we introduced a 0.2~MHz frequency offset between the AOMs. The blue curve of Fig.~\ref{fig:spectrum} (upper panel) shows the measured spectrum when no modulation is applied. Here two distinct peaks can be observed at $\SI{2.2}{ \mega \hertz}$ and $\SI{2.4}{\mega \hertz}$. These correspond to the beat signal from each of the sites, A (the desired signal, containing the information of the phase noise in the fiber) and B (the undesired signal, resulting from other users in the network). The red curve shows the equivalent spectrum when applying a frequency hopping sequence with eight hopping frequencies. While the desired signal resulting from remote A remains almost unaltered, the undesired signal is spread across a large bandwidth and strongly suppressed.

According to Eq.~(\ref{eq:PSD_0}), this suppression becomes stronger for increased number of hopping levels. In Fig.~\ref{fig:spectrum} (lower panel) we show the variation of the relative height between the two signals with increasing number of hopping frequencies (red circles) and compare it with the prediction from Eq.~(\ref{eq:PSD_0}) (red dashed line). Here we have normalized both signals to 1 at ${\nhopping} = 1$. While our experimental data shows a slightly weaker suppression than the one predicted by Eq.~(\ref{eq:PSD_0}), both theory and data show remarkable agreement. The solid line shows a fit to the data following $1/(\nhopping)^{3.5 \pm 0.3}$. Importantly, we demonstrate an attenuation of the undesired signal resulting from the other branche by about $\SI{35}{\deci\bel}$ using eight hopping levels.


\section{Conclusion}
We have demonstrated a code-division multiple access technique using frequency hopping spread spectrum to disseminate an ultrastable optical frequency to independent users in a branching passive optical network. By spreading and despreading the spectrum we suppressed undesired signals resulting from other users in the network. We provide a quantitative measure of the suppression of the signals from other nodes, which scales as $\nhopping^{-4}$. Using this technique we demonstrate ultrastable optical frequency transfer, indicating the possibility of relative frequency stability well below $10^{-18}$ at $\SI{100}{\second}$ of averaging time while having about 144 users operating in the same network. The demonstrated stability does not suffer from any fundamental limitation, and can be straightforwardly improved for applications where higher performance is required. In the future, the demonstrated technique could be used to simplify the operational requirements of time-frequency distribution networks. We furthermore point out the possibility of encrypting ultrastable optical signals using CDMA at the ultrastable laser source itself.

\section{Supplementary materials}

\subsection{Power spectrum calculation}
\label{s:FHSSdev}
In the following, we will make a number of simplifying assumptions. First, we assume the propagation delay in both fiber links to be the same, i.e. $\tau_{B} = \tau_{A}$. If $\tau_{B} \neq \tau_{A}$, an additional frequency modulation will occur. This will result in a stronger suppression of the DC component shown in Eq.~\ref{eq:PSD_0}. Here we assume the less optimal condition where both codes have aligned symbol transitions and maintain a fixed time offset due to the fact that $\tau_{B} = \tau_{A}$. We consider the case where the programmable delay $\tau_\mathrm{d}$ has been correctly determined by experimental means (see section~\ref{s:FDC} below) and set to $\tau_\mathrm{d}=\tau_\mathrm{A}$. We also assume the second remote location to be locked, such that $2\pi t\times \delta\!f_{\mathrm{B}} + \delta\!\phi_{B} \approx 0$. Finally, we assume that the hopping sequence is periodic in time, with repetition time $T$. This condition is also met in our experiment.


Under these assumptions, the output signal of the mixer that follows from the interference of branch A and B [Eq.~(\ref{eq:SmixerB})] can be expressed as
\begin{eqnarray}
\label{time_dependent}
 V_{\mathrm{mixer,AB}}(t) &=&  V_\mathrm{0,AB} \sin (4\pi \Delta\!f\left[s_\mathrm{B}(t-2 \tau_\mathrm{A}) - s_\mathrm{A}(t-2\tau_{\mathrm{A}})\right] t + \phi_\mathrm{disp}(t)) \nonumber \\ 
 &=& V_\mathrm{0,AB} \sum_{n=1}^{\nbits} \Theta_n(t) \sin (4\pi\Delta\!f m_{n}t +\phi_{n}),\nonumber\\
\end{eqnarray}
where $V_\mathrm{0,AB}$ is the signal amplitude, and the residual (unsuppressed) phase modulation due to dispersion is: 
\begin{eqnarray}
\label{phi_d}
    \phi_\mathrm{disp}(t) &=& 2\pi \Disp{\mathrm{AB}} \Delta f \cdot s_\mathrm{B}(t-2\tau_\mathrm{A}) -\Delta \varphi\cdot s_\mathrm{A} (t-2\tau_{\mathrm{A}})  + \varphi_\mathrm{slow}.
\end{eqnarray}
Here, the function $\Theta_n(t)$ takes the value of 1 in the interval $[n t_{s}, (n+1) t_{s}]$, and 0 elsewhere. Furthermore, $t_{s}$ is the symbol duration, $m_{n}=s_\mathrm{B}(n t_{s}-2 \tau_\mathrm{A}) - s_\mathrm{A}(n t_{s}-2\tau_{\mathrm{A}})$ is the difference between the two pseudo-random sequences, which is constant in the interval $[n t_{s}, (n+1) t_{s}]$, and $\nbits$ is the total number of symbols in the sequence (so that the duration of the pseudo-random sequence becomes $\nbits t_\mathrm{s}=T$).  Finally $\phi_{n}$ represents the evaluation of Eq.~(\ref{phi_d}) over the interval $[n t_{s}, (n+1) t_{s}]$. Within this interval, $\phi_{n}$ is constant, and in general $\phi_{n}$ can take $N_{\mathrm{hopping}}^{2}$ possible values.

Because Eq.~(\ref{time_dependent}) repeats itself after a time $T$, it can be expressed in terms of its Fourier series as follows:
\begin{align}
\label{eq:fourier}
V_{\mathrm{mixer,AB}}(t) =\sum_{k=-\infty}^{\infty} C_{k}e^{i\omega_{k} t}, && C_{k} =\frac{1}{T} \int_{0}^{T} V_{\mathrm{mixer,AB}}(t)e^{-i\omega_{k} t},
\end{align}
where $\omega_{k} = 2\pi k/T$. In what follows, we will make use of the simplified expression $4 \pi\,\Delta\!f\!\,m_{n}\equiv \omega_{n}$. For simplicity, we will also choose $V_{\mathrm{0,AB}}=1$. The Fourier coefficients can then be calculated using Eqs.~(\ref{eq:fourier}) and (\ref{time_dependent}):
\begin{multline}
\label{fourier_serie2s}
C_{k} = \frac{1}{T}\int_{0}^T \sum_{n=1}^{\nbits}\Theta_n(t) \sin (\omega_{n} t  +\phi_{n})e^{-i\omega_{k}t} = \frac{1}{T} \sum_{n=1}^{\nbits}\int_{n t_{s}}^{ (n+1) t_{s}} \sin (\omega_{n} t  +\phi_{n})e^{-i\omega_{k}t}.
\end{multline}

The righthand side of Eq.~(\ref{fourier_serie2s}) can be straightforwardly evaluated to obtain:

\begin{eqnarray}
\label{C:K}
 C_{k}&=& \frac{1}{2T}\sum_{n=1}^{\nbits} \left( e^{i \phi_{n}} \frac{e^{i (\omega_{n} - \omega_{k})n t_{s}} - e^{i(\omega_{n} - \omega_{k})(n+1) t_{s}} }{\omega_{n} - \omega_{k}} \right. \nonumber \\
&& \left. +\; e^{-i\phi_{n}}\frac{e^{-i (\omega_{n} + \omega_{k})n t_{s}} - e^{-i (\omega_{n} + \omega_{k})(n+1) t_{s}} }{\omega_{n} + \omega_{k}} \right) .
\end{eqnarray}

The DC frequency component of $V_{\mathrm{mixer,AB}}(t)$ is given by the $k=0$ Fourier coefficient $C_0$, which according to Eq.~(\ref{C:K}) becomes
\begin{equation}
\label{fourier_serie4s}
C_{0} = \frac{1}{T}\sum_{n=1}^{\nbits} \frac{\cos({\omega_{n} n t_{s}}  +\phi_{n})-\cos({\omega_{n} (n+1)t_{s}} +\phi_{n})}{\omega_{n}}, 
\end{equation}

This expression is dominated by the terms for which $\omega_n=0$. Evaluating the summand in the limit $\omega_{n} \to 0$  and keeping only the terms where the condition $\omega_n=0$ is fulfilled gives:

\begin{equation}
\label{eq:limfi}
\lim_{\omega_{n} \to 0} C_{0} = \frac{t_{s}}{T}\sum_{n=1}^{\nbits }\sin{\phi_{n}}\cdot\delta_{\omega_{n},0}. 
\end{equation}

If the number of symbols is much larger than the number of hopping levels (i.e., $\nbits \gg \nhopping $) the total number of non-zero values of Eq.~(\ref{eq:limfi}) can be approximated to be $ \lfloor\nbits / \nhopping \rfloor $. Restricting the previous summation to the non-zero values, Eq.~(\ref{eq:limfi}) becomes:
%
\begin{equation}
    \lim_{\omega_{n} \to 0} C_{0} 
 \approx  \frac{t_{s}}{T}\sum_{n=1}^{\lfloor \nbits/\nhopping \rfloor }\sin{\phi^{\prime}_{n}}, 
\end{equation}
where $\phi^{\prime}_{n}$ indicates that the summation runs over the subset of $\phi_{n}$ values where the condition $\omega_{n} =0$ [implemented via the Kronecker delta function in Eq.~(\ref{eq:limfi})] is fulfilled.

To further simplify the previous equation we note that the probability of having identical PRFHS symbols in branches A and B is $1/\nhopping$ so each of the $\phi^{\prime}_{n}$ values will appear with an average count of $\nbits/\nhopping^2$ within the previous summation:
\begin{eqnarray}
    \label{psd_finals}
  \lim_{\omega_{n} \to 0} C_{0}  &&\approx\frac{t_{s}}{T}\sum_{n=1}^{\lfloor \nbits/\nhopping \rfloor }\sin{\phi^{\prime}_{n}} \nonumber  \\
  &&\approx \frac{ t_{s}}{T}\sum_{n = 1}^{\nhopping}\frac{\nbits}{\nhopping^2}\sin{\phi^{\prime}_{n,\mathrm{unique}}}
 \approx \frac{\sum_{n = 1}^{\nhopping}\sin{\phi^{\prime}_{n,\mathrm{unique}}}}{ \nhopping ^2},
\end{eqnarray}
where we have used $T = t_{s}\cdot\nbits$, and $\phi^{\prime}_{n,\mathrm{unique}}$ indicates that the summation runs only over the set of unique $\phi^{\prime}_{n}$ values.  Finally, the power spectral density follows from taking the square of Eq.~(\ref{psd_finals}).

Equation~(\ref{psd_finals}) reveals the presence of an additional 'cross talk' phase offset, proportional to $1/\nhopping^{2}$. This phase offset is quasi-static: it will vary slowly as the dispersion changes slowly over time, primarily due to temperature variations in the fiber and varying delays in sections of uncompensated fiber in locations A and B. As such, it will cause some residual frequency instability, which can be reduced by increasing $\nhopping$. 
\subsection{Calibration procedure}
\label{s:FDC}

A crucial aspect of our protocol is the proper configuration of the demodulation signal applied to the mixer such that:
\begin{align}
\tau_{\mathrm{d}} &= \tau &  \Delta \varphi = \frac{\partial \varphi}{ \partial \omega} \cdot 2\pi\Delta\!f 
\end{align}
%
By setting $\tau_{\mathrm{d}} = \tau $ and $ \Delta \varphi = \frac{\partial \varphi}{ \partial \omega} \cdot 2\pi\Delta\!f  $, 
a cancellation of terms in Eq.~(\ref{eq:Smixer_With_phase_delays}) occurs such that $\phi_\mathrm{mixer}(t) =  2\left[2 \pi t \delta\!f + \delta\!\phi(t)\right]$, which can be used to cancel the link noise as before. Here, we describe how the above conditions are achieved in practice. 
We can relate the spectrum measured at each remote location to both $\tau$ and $\varphi$, allowing us to achieve the above optimal condition. To see this, we recall Eq.~(\ref{eq:Smixer_With_phase_delays}):
\begin{eqnarray}
        \phi_\mathrm{mixer}(t) &=& 2\pi t \times \left[2\Delta\!f(s(t-2\tau) - s(t-2\tau_{\mathrm{d}}))\right]  + 2\left[2 \pi t \delta\!f + \delta\!\phi(t)\right] \nonumber \\  &+& \frac{\partial \varphi}{ \partial \omega} \cdot 2\pi\Delta\!f \cdot s(t-2\tau) - \Delta \varphi\cdot s(t-2\tau_{\mathrm{d}}) .
\end{eqnarray}
\begin{figure}[t!]
\centering\includegraphics[width= \textwidth]{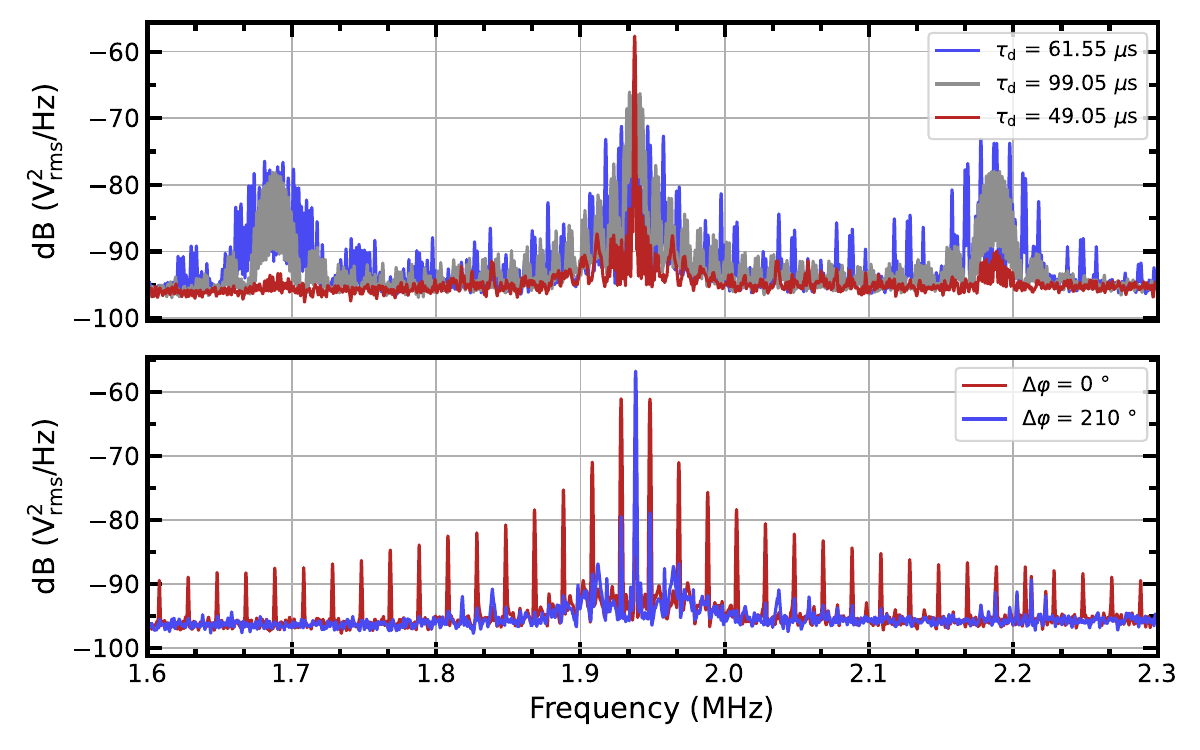}
\caption{(Upper panel) The data shows the PSD for three different delays measured after the second mixing stage. When $\tau_{\mathrm{d}}$ is not properly configured an additional modulation occurs at $2\Delta\!f$ (grey and blue curves). If this delay matches the transmission delay on the fiber the modulation disappears (red curve). Here, the PRFHS parameters are $t_{s} = \SI{50}{\micro \second}$, $ \nhopping = 2$, $ \Delta\!f = \SI{0.125}{\mega \hertz}$ and $\nbits = 7$. (Lower panel) PSD when the effect of optical dispersion is present (red curve) or when it is eliminated (blue curve). Here our sequence simply alternates between two hopping levels with paremeters $t_{s} = \SI{2}{\micro \second}$, $ \Delta\!f = \SI{3}{\mega \hertz}$}
\label{fig:spectrum_sup}
\end{figure}
Note that if $\tau_{\mathrm{d}} \neq \tau $, a modulation with frequencies at integer multiples of $2\Delta\!f$ will occur. By minimizing the power spectral density (PSD) at these undesired frequencies we can achieve the condition $\tau_{\mathrm{d}} = \tau$. In Fig.~\ref{fig:spectrum_sup} (upper panel) we show the PSD for the case where the two-bit sequences are time-aligned (red curve), as well as for cases where they are misaligned by $t_{s}/2 = \SI{12.5}{\micro \second}$ (blue curve) and by $2t_{s} = \SI{50}{ \micro \second}$ (grey curve). In our case, by minimizing the height of these undesired peaks we obtain an optimal delay of $\tau_{\mathrm{d}} = \SI{49.05}{\micro \second}$ (in agreement with the expected delay for a fiber spool with a length of about $\SI{10}{\kilo \meter}$). Since the strength of this suppression becomes larger for smaller $t_{s}$, we start with a large value of $t_{s} = \SI{50}{\micro \second}$ (so the peaks are clearly visible) and progressively reduce it to its operational value ($t_{s} =\SI{2}{\micro \second}$).

We can employ a similar procedure to set the optimal values of $\Delta\varphi$. Note that, the phase jumps occur at a rate of $1/t_{s}$, which is reflected by the PSD in the form of a series of harmonics. Similarly as before, we can set the optimal $\Delta\varphi_\mathrm{mixer} $ by measuring the modulation peak in the PSD and minimizing it. Figure \ref{fig:spectrum_sup} (lower panel) shows the PSD when $\Delta\varphi_\mathrm{mixer} = 0$ and when this additional modulation is eliminated with $\Delta\varphi/\Delta{f}= \SI[per-mode=repeated-symbol]{70}{\degree \per \mega \hertz}$.

\begin{backmatter}
\bmsection{Funding}
We acknowledge support from the Dutch Ministry of Economic Affairs and Climate Policy (EZK), as part of the Quantum Delta NL program.

\bmsection{Acknowledgments}
We thank Kjeld Eikema for providing the ultrastable laser used in this work.  
\bmsection{Disclosures}
J.C.J.K. is co-founder and shareholder of OPNT bv, a company in network time and frequency distribution. The research findings presented here are not related to any products or services currently provided by OPNT bv. The authors declare no further conflicts of interest.

\bmsection{Data availability} 
Data underlying the results presented in this paper are not publicly available at this time but may be obtained from the authors upon reasonable request.

\end{backmatter}


\bibliography{Optica-template}

\begin{thebibliography}{10}
\newcommand{\enquote}[1]{``#1''}

\bibitem{Maol1990}
L.-S. Ma, P.~Jungner, J.~Ye, and J.~L. Hall, \enquote{Delivering the same
  optical frequency at two places: accurate cancellation of phase noise
  introduced by an optical fiber or other time-varying path,}
  {\protect\JournalTitle{Opt. Lett.}} \textbf{19}, 1777--1779 (1994).

\bibitem{WilliamsJOSAB2008}
P.~A. Williams, W.~C. Swann, and N.~R. Newbury, \enquote{High-stability
  transfer of an optical frequency over long fiber-optic links,}
  {\protect\JournalTitle{J. Opt. Soc. Am. B}} \textbf{25}, 1284--1293 (2008).

\bibitem{PredehlScience2012}
K.~Predehl, G.~Grosche, S.~M.~F. Raupach, S.~Droste, O.~Terra, J.~Alnis,
  T.~Legero, T.~W. H{\"a}nsch, T.~Udem, R.~Holzwarth, and H.~Schnatz,
  \enquote{A 920-kilometer optical fiber link for frequency metrology at the
  19th decimal place,} {\protect\JournalTitle{Science}} \textbf{336}, 441--444
  (2012).

\bibitem{DrostePrl2013}
S.~Droste, F.~Ozimek, T.~Udem, K.~Predehl, T.~W. H\"ansch, H.~Schnatz,
  G.~Grosche, and R.~Holzwarth, \enquote{Optical-frequency transfer over a
  single-span 1840 km fiber link,} {\protect\JournalTitle{Phys. Rev. Lett.}}
  \textbf{111}, 110801 (2013).

\bibitem{Schioppo2022natcoms}
M.~Schioppo, J.~Kronj{\"a}ger, A.~Silva, R.~Ilieva, J.~W. Paterson, C.~F.~A.
  Baynham, W.~Bowden, I.~R. Hill, R.~Hobson, A.~Vianello,
  M.~Dovale-{\'A}lvarez, R.~A. Williams, G.~Marra, H.~S. Margolis,
  A.~Amy-Klein, O.~Lopez, E.~Cantin, H.~{\'A}lvarez-Mart{\'\i}nez,
  R.~Le~Targat, P.~E. Pottie, N.~Quintin, T.~Legero, S.~H{\"a}fner, U.~Sterr,
  R.~Schwarz, S.~D{\"o}rscher, C.~Lisdat, S.~Koke, A.~Kuhl, T.~Waterholter,
  E.~Benkler, and G.~Grosche, \enquote{Comparing ultrastable lasers at 7
  $\times$ 10-17 fractional frequency instability through a 2220 km optical
  fibre network,} {\protect\JournalTitle{Nat. Commun.}} \textbf{13}, 212
  (2022).

\bibitem{clonets}
\url{https://clonets-ds.eu/}.

\bibitem{CantinNJP2021}
E.~Cantin, M.~Tønnes, R.~L. Targat, A.~Amy-Klein, O.~Lopez, and P.-E. Pottie,
  \enquote{An accurate and robust metrological network for coherent optical
  frequency dissemination,} {\protect\JournalTitle{New Journal of Physics}}
  \textbf{23}, 053027 (2021).

\bibitem{HusmannOE2021}
D.~Husmann, L.-G. Bernier, M.~Bertrand, D.~Calonico, K.~Chaloulos, G.~Clausen,
  C.~Clivati, J.~Faist, E.~Heiri, U.~Hollenstein, A.~Johnson, F.~Mauchle,
  Z.~Meir, F.~Merkt, A.~Mura, G.~Scalari, S.~Scheidegger, H.~Schmutz,
  M.~Sinhal, S.~Willitsch, and J.~Morel, \enquote{Si-traceable frequency
  dissemination at 1572.06\&\#x00a0; nm in a stabilized fiber network with ring
  topology,} {\protect\JournalTitle{Opt. Express}} \textbf{29}, 24592--24605
  (2021).

\bibitem{InseroScientificReports2017}
G.~Insero, S.~Borri, D.~Calonico, P.~C. Pastor, C.~Clivati, D.~D'Ambrosio,
  P.~De~Natale, M.~Inguscio, F.~Levi, and G.~Santambrogio, \enquote{Measuring
  molecular frequencies in the 1--10{\thinspace}$\mu$m range at 11-digits
  accuracy,} {\protect\JournalTitle{Scientific Reports}} \textbf{7}, 12780
  (2017).

\bibitem{ClivatiNature2022}
C.~Clivati, A.~Meda, S.~Donadello, S.~Virzi, M.~Genovese, F.~Levi, A.~Mura,
  M.~Pittaluga, Z.~Yuan, A.~J. Shields, M.~Lucamarini, I.~P. Degiovanni, and
  D.~Calonico, \enquote{Coherent phase transfer for real-world twin-field
  quantum key distribution,} {\protect\JournalTitle{NATURE COMMUNICATIONS}}
  \textbf{13} (2022).

\bibitem{LucamarininiNature2018}
M.~Lucamarini, Z.~L. Yuan, J.~F. Dynes, and A.~J. Shields, \enquote{Overcoming
  the rate--distance limit of quantum key distribution without quantum
  repeaters,} {\protect\JournalTitle{Nature}} \textbf{557}, 400--403 (2018).

\bibitem{MarraScience2018}
G.~Marra, C.~Clivati, R.~Luckett, A.~Tampellini, J.~Kronjäger, L.~Wright,
  A.~Mura, F.~Levi, S.~Robinson, A.~Xuereb, B.~Baptie, and D.~Calonico,
  \enquote{Ultrastable laser interferometry for earthquake detection with
  terrestrial and submarine cables,} {\protect\JournalTitle{Science}}
  \textbf{361}, 486--490 (2018).

\bibitem{MarraScience2022}
G.~Marra, D.~M. Fairweather, V.~Kamalov, P.~Gaynor, M.~Cantono, S.~Mulholland,
  B.~Baptie, J.~C. Castellanos, G.~Vagenas, J.-O. Gaudron, J.~Kronjäger, I.~R.
  Hill, M.~Schioppo, I.~B. Edreira, K.~A. Burrows, C.~Clivati, D.~Calonico, and
  A.~Curtis, \enquote{Optical interferometry–based array of seafloor
  environmental sensors using a transoceanic submarine cable,}
  {\protect\JournalTitle{Science}} \textbf{376}, 874--879 (2022).

\bibitem{MatveevPRL2013}
A.~Matveev, C.~G. Parthey, K.~Predehl, J.~Alnis, A.~Beyer, R.~Holzwarth,
  T.~Udem, T.~Wilken, N.~Kolachevsky, M.~Abgrall, D.~Rovera, C.~Salomon,
  P.~Laurent, G.~Grosche, O.~Terra, T.~Legero, H.~Schnatz, S.~Weyers,
  B.~Altschul, and T.~W. H\"ansch, \enquote{Precision measurement of the
  hydrogen $1s\mathrm{\text{\ensuremath{-}}}2s$ frequency via a 920-km fiber
  link,} {\protect\JournalTitle{Phys. Rev. Lett.}} \textbf{110}, 230801 (2013).

\bibitem{ChouScience2010}
C.~W. Chou, D.~B. Hume, T.~Rosenband, and D.~J. Wineland, \enquote{Optical
  clocks and relativity,} {\protect\JournalTitle{Science}} \textbf{329},
  1630--1633 (2010).

\bibitem{GrottiNatPhys2018}
J.~Grotti, S.~Koller, S.~Vogt, S.~Haefner, U.~Sterr, C.~Lisdat, H.~Denker,
  C.~Voigt, L.~Timmen, A.~Rolland, F.~N. Baynes, H.~S. Margolis, M.~Zampaolo,
  P.~Thoumany, M.~Pizzocaro, B.~Rauf, F.~Bregolin, A.~Tampellini, P.~Barbieri,
  M.~Zucco, G.~A. Costanzo, C.~Clivati, F.~Levi, and D.~Calonico,
  \enquote{Geodesy and metrology with a transportable optical clock,}
  {\protect\JournalTitle{NATURE PHYSICS}} \textbf{14}, 437+ (2018).

\bibitem{TakamotoNatPhot2020}
M.~Takamoto, I.~Ushijima, N.~Ohmae, T.~Yahagi, K.~Kokado, H.~Shinkai, and
  H.~Katori, \enquote{Test of general relativity by a pair of transportable
  optical lattice clocks,} {\protect\JournalTitle{NATURE PHOTONICS}}
  \textbf{14}, 411+ (2020).

\bibitem{Guillou-CamargoAppliedOptics2018}
F.~Guillou-Camargo, V.~M\'{e}noret, E.~Cantin, O.~Lopez, N.~Quintin,
  E.~Camisard, V.~Salmon, J.-M.~L. Merdy, G.~Santarelli, A.~Amy-Klein, P.-E.
  Pottie, B.~Desruelle, and C.~Chardonnet, \enquote{First industrial-grade
  coherent fiber link for optical frequency standard dissemination,}
  {\protect\JournalTitle{Appl. Opt.}} \textbf{57}, 7203--7210 (2018).

\bibitem{GroscheOL2014}
G.~Grosche, \enquote{Eavesdropping time and frequency: phase noise cancellation
  along a time-varying path, such as an optical fiber,}
  {\protect\JournalTitle{Opt. Lett.}} \textbf{39}, 2545--2548 (2014).

\bibitem{BaiOL2013}
Y.~Bai, B.~Wang, X.~Zhu, C.~Gao, J.~Miao, and L.~J. Wang, \enquote{Fiber-based
  multiple-access optical frequency dissemination,} {\protect\JournalTitle{Opt.
  Lett.}} \textbf{38}, 3333--3335 (2013).

\bibitem{SchediwyOE2010}
S.~W. Schediwy, D.~Gozzard, K.~G.~H. Baldwin, B.~J. Orr, R.~B. Warrington,
  G.~Aben, and A.~N. Luiten, \enquote{High-precision optical-frequency
  dissemination on branching optical-fiber networks,}
  {\protect\JournalTitle{Opt. Lett.}} \textbf{38}, 2893--2896 (2013).

\bibitem{GramiBook2016}
A.~Grami, \enquote{Communication networks,} in \emph{Introduction to Digital
  Communications,}  (Elsevier, 2016), pp. 457--491.

\bibitem{Jochen_pattent}
J.~Kronjäger, \enquote{Frequency and/or time transfer apparatus, system and
  method,}  (European Patent EP4289088A1, March. 2022).

\bibitem{XieTUFFC2024}
K.~Xie, X.~Zhang, L.~Hu, J.~Chen, and G.~Wu, \enquote{Fiber-optic time transfer
  based on bidirectional fdm and cross correlation processing,}
  {\protect\JournalTitle{IEEE Transactions on Instrumentation and Measurement}}
  \textbf{73}, 1--7 (2024).

\bibitem{Sotiropoulos2013}
N.~Sotiropoulos, C.~M. Okonkwo, R.~Nuijts, H.~de~Waardt, and J.~C.~J.
  Koelemeij, \enquote{Delivering 10 gb/s optical data with picosecond timing
  uncertainty over 75 km distance,} {\protect\JournalTitle{Opt. Express}}
  \textbf{21}, 32643--32654 (2013).

\bibitem{PinkertApplOpt2015}
T.~J. Pinkert, O.~B\"{o}ll, L.~Willmann, G.~S.~M. Jansen, E.~A. Dijck, B.~G.
  H.~M. Groeneveld, R.~Smets, F.~C. Bosveld, W.~Ubachs, K.~Jungmann, K.~S.~E.
  Eikema, and J.~C.~J. Koelemeij, \enquote{Effect of soil temperature on
  optical frequency transfer through unidirectional
  dense-wavelength-division-multiplexing fiber-optic links,}
  {\protect\JournalTitle{Appl. Opt.}} \textbf{54}, 728--738 (2015).

\bibitem{Ghosh1994}
G.~Ghosh, M.~Endo, and T.~Iwasaki, \enquote{Temperature-dependent sellmeier
  coefficients and chromatic dispersions for some optical fiber glasses,}
  {\protect\JournalTitle{Journal of Lightwave Technology}} \textbf{12},
  1338--1342 (1994).

\bibitem{pyrpl}
L.~Neuhaus, R.~Metzdorff, S.~Chua, T.~Jacqmin, T.~Briant, A.~Heidmann, P.-F.
  Cohadon, and S.~Deléglise, \enquote{Pyrpl (python red pitaya lockbox) — an
  open-source software package for fpga-controlled quantum optics experiments,}
  in \emph{2017 Conference on Lasers and Electro-Optics Europe \& European
  Quantum Electronics Conference (CLEO/Europe-EQEC),}  (2017), pp. 1--1.

\end{thebibliography}






\end{document}